%% file: 1.5T1Fe.tex
\documentclass[conference]{IEEEtran}
\pdfoutput=1
\usepackage{cite}
\usepackage{amsmath,amssymb,amsfonts}
\usepackage{algorithmic}
\usepackage{graphicx}
\usepackage{textcomp}
\usepackage{xcolor}
\usepackage{csquotes}
\usepackage{multirow}
\usepackage{booktabs}
\usepackage{colortbl}
\definecolor{Gray}{gray}{0.94}

\newcommand{\textoverline}[1]{\ensuremath{\overline{\text{#1}}}}
\usepackage[a4paper, total={184mm,239mm}]{geometry}
\usepackage{comment}
\usepackage{xspace}
\usepackage[acronyms]{glossaries}
\input{glossar.tex}

\usepackage{array}
\newcolumntype{?}{!{\vrule width 1pt}}

\def\BibTeX{{\rm B\kern-.05em{\sc i\kern-.025em b}\kern-.08em
    T\kern-.1667em\lower.7ex\hbox{E}\kern-.125emX}}

\usepackage{subcaption}
\usepackage{gensymb}
\usepackage{cleveref}
\crefname{enumi}{Step}{Steps}
\crefname{section}{Sec.}{Sec.}
\crefname{subsection}{Sec.}{Sec.}
\crefname{figure}{Fig.}{Fig.}
\crefname{algocf}{Algorithm}{Algorithms}
\crefname{algorithm}{Algorithm}{Algorithms}
\crefname{algocf}{Algorithm}{Algorithms}
\crefname{table}{Tab.}{Tab.}
\crefname{equation}{Eq.}{Eq.}
\crefname{eqnarray}{Eq.}{Eq.}
\crefname{appendix}{Section}{Sections}

\begin{document}

\bstctlcite{IEEEexample:BSTcontrol}

\title{Compact and High-Performance TCAM \\ Based on Scaled Double-Gate FeFETs\\
\vspace{-1ex}
}

\author{Liu~Liu\textsuperscript{$\dagger$}, Shubham~Kumar\textsuperscript{$\star \ddag$}, 
Simon~Thomann\textsuperscript{$\star$}, Yogesh Singh Chauhan\textsuperscript{$\ddag$}, Hussam Amrouch\textsuperscript{$\S$}, and Xiaobo Sharon Hu\textsuperscript{$\dagger$} \\
\textsuperscript{$\dagger$}Department of Computer Science and Engineering, 
University of Notre Dame, USA, \\\textsuperscript{$\star$}Chair of Semiconductor Test and Reliability (STAR),
University of Stuttgart, Germany 
\\\textsuperscript{$\ddag$}Department of Electrical Engineering, Indian Institute of Technology Kanpur, India
\\\textsuperscript{$\S$}Chair of AI Processor Design,
Technical University of Munich (TUM), Germany
\\
Email: \{lliu24, shu\}@nd.edu, amrouch@tum.de
\vspace{-3ex}}

\maketitle

\thispagestyle{empty}

\begin{abstract}

Ternary content addressable memory (TCAM), widely used in network routers and high-associativity caches, is gaining popularity in machine learning and data-analytic applications.  Ferroelectric FETs (FeFETs) are a promising candidate for implementing TCAM owing to their high ON/OFF ratio, non-volatility, and CMOS compatibility. However, conventional single-gate FeFETs (SG-FeFETs) suffer from relatively high write voltage, low endurance, potential read disturbance, and face scaling challenges. Recently, a double-gate FeFET (DG-FeFET) has been proposed and outperforms SG-FeFETs in many aspects. This paper investigates TCAM design challenges specific to DG-FeFETs and introduces a novel 1.5T1Fe TCAM design based on DG-FeFETs. A 2-step search with early termination is employed to reduce the cell area and improve energy efficiency. A shared driver design is proposed to reduce the peripherals area. Detailed analysis and SPICE simulation show that the 1.5T1Fe DG-TCAM leads to superior search speed and energy efficiency. The 1.5T1Fe TCAM design can also be built with SG-FeFETs, which achieve search latency and energy improvement compared with 2FeFET TCAM. 

\end{abstract}


\section{Introduction}
\label{introduction}

Ternary content addressable memories (TCAMs) support fast parallel search directly in the memory, which is a promising in-memory-computing (IMC) kernel to address the processor-memory bottleneck~\cite{hu_iedm21}. Beyond conventional applications, such as network routers and associative caches, TCAMs have been used for a variety of data-centric applications, such as machine learning, neuromorphic computing, and bioinformatics~\cite{ann_iccad20,Liu2023mhcam,pedretti_nc21,ni_ne19,li_ted21,searchd_tcad19}.  

The design space for content addressable memory (CAM) has been rapidly expanding in recent years largely due to the employment of non-volatile memories (NVMs) in CAM design. Unlike the conventional CMOS TCAM with large area and high energy overhead, NVM-based TCAMs (NV-TCAMs) are more compact, energy-efficient, and non-volatile~\cite{nvcam_survey_2015}, which are more suitable for many emerging applications. A wide range of NV-TCAM designs have been proposed based on two-terminal NVMs, such as resistive RAM (RRAM)~\cite{li_ted21,2.5t1rram_isscc16,3t1rram_isscc15}, phase change memory (PCM)~\cite{2t2pcm_jssc14}, spin transfer torque magnetic RAM (STT-MRAM)~\cite{stt-mram_tcas19}, and three-terminal ferroelectric FETs (FeFETs)~\cite{yin_tcas18,1t2fe_date21}. The two-terminal NVM-based TCAMs typically require current-driven write schemes and large access transistors, leading to higher energy consumption. The low ON/OFF ratio of many two-terminal NVMs often requires more transistors in the CAM cell design, limits the word length of the TCAM array, and makes sensing more challenging~\cite{2.5t1rram_isscc16,2t2pcm_jssc14,eva-cam}. The three-terminal FeFET is a promising candidate to implement NV-TCAMs for its high ON/OFF ratio, high OFF resistance, relatively low write energy, and CMOS compatibility~\cite{28fefet_iedm16}. Recent work proposed to use only two FeFETs to build an ultra-high density TCAM cell~\cite{yin_tcas18}.  

\textbf{However, conventional FeFETs also face key challenges.} \textbf{(1)} The thick ferroelectric (FE) layer ($\sim$10 nm) incurs severe charge trapping which limits the endurance, read throughput, and reliability of FeFET~\cite{dg-fefet_vlsi21,tan2021ferroelectric}. \textbf{(2)} Although the electric field-driven write scheme of FeFETs is energy efficient, it requires $\pm$4V write voltage to switch the state of the FE layer. The write voltage is higher than most standard CMOS technologies hence posing challenges to the write drivers and high-voltage tolerance for other peripherals. For emerging applications with seldom writes and frequent searches, the write drivers stay idle most time but consume a large area and high leakage power. \textbf{(3)} Conventional FeFETs write and read the state from the same gate of a FeFET, which may lead to the read disturbance issue. 

Recently, a double-gate FeFET (DG-FeFET) device has been proposed to mitigate the high write voltage and charge trapping concerns~\cite{dg-fefet_vlsi21}. By reducing the FE thickness and using the separated front gate (FG) and back gate (BG) for write and read operation, respectively, DG-FeFETs only require 2V write/read voltage and facilitate technology scaling~\cite{reliability_ted22}. The separated write/read path avoids accumulated read disturbance presented in conventional FeFETs. The lower write voltage also helps to improve the endurance of the FeFET ($>\!\!10^{10}$)~\cite{tan2021ferroelectric}. Nevertheless, DG-FeFETs need BGs with separate control during read, resulting in area penalties. We refer to the conventional FeFETs as single-gate FeFETs (SG-FeFETs) to be differentiated from the emerging DG-FeFETs. 

It comes naturally to consider implementing the 2FeFET TCAM design based on DG-FeFET. However, the 2 FeFETs per cell design magnifies the BG control overhead. This is also a common concern that NV-TCAM cells are typically implemented by two NVM devices with several transistors. In general, NVM devices are more expensive than CMOS transistors and increase the overhead of peripherals. Single NVM per TCAM cell design can be a possible solution. 



This paper introduces a compact and high-performance TCAM design with only one FeFET per cell. Our contributions are below.
\begin{itemize}
    \item We propose a 1.5T1Fe TCAM design using only one DG-FeFET per cell, which minimizes the BG control overhead and achieves fast parallel search. A two-step search with an early termination scheme is adopted to reduce the cell area and achieve superior energy efficiency. Simulation results show that our proposed 1.5T1DG-Fe TCAM design achieves 4$\times$/1.21$\times$/1.21$\times$ write energy/search latency/search energy improvement compared to 2SG-FeFET TCAM, and 1.83$\times$/3.79$\times$ cell area/search energy improvement compared to 16T CMOS TCAM. To the best of our knowledge, this is the first DG-FeFET-based CAM design. 
    \item We also build the widely-adopted 2FeFET TCAM based on DG-FeFET and find that the 2DG-FeFET TCAM cannot achieve competitive performance due to the DG-FeFET device limitation.
    \item  We introduce a shared driver design across the TCAM subarrays through device-circuit co-optimization to reduce driver area and improve driver utilization.
    \item The proposed 1.5T1Fe TCAM design can also be implemented by SG-FeFET and achieve 2$\times$/1.66$\times$/1.42$\times$ write energy/search latency/search energy improvement compared to 2SG-FeFET TCAM, and 2.12$\times$/4.42$\times$ cell area and search energy improvement compared to CMOS TCAM. 
\end{itemize}



\section{Preliminaries}
\label{preliminaries}


\subsection{Single-Gate \& Double-Gate FeFET}
\label{dg-fefet}

\begin{figure}[t]
    \centering
    \begin{subfigure}{\columnwidth}
        \phantomsubcaption
        \label{fig:single_gate}
        \phantomsubcaption
        \label{fig:double_gate}
        \phantomsubcaption
        \label{fig:SG_idvg}
        \phantomsubcaption
        \label{fig:DG_idvg}
    \end{subfigure}
    \includegraphics[ width=0.8\linewidth] {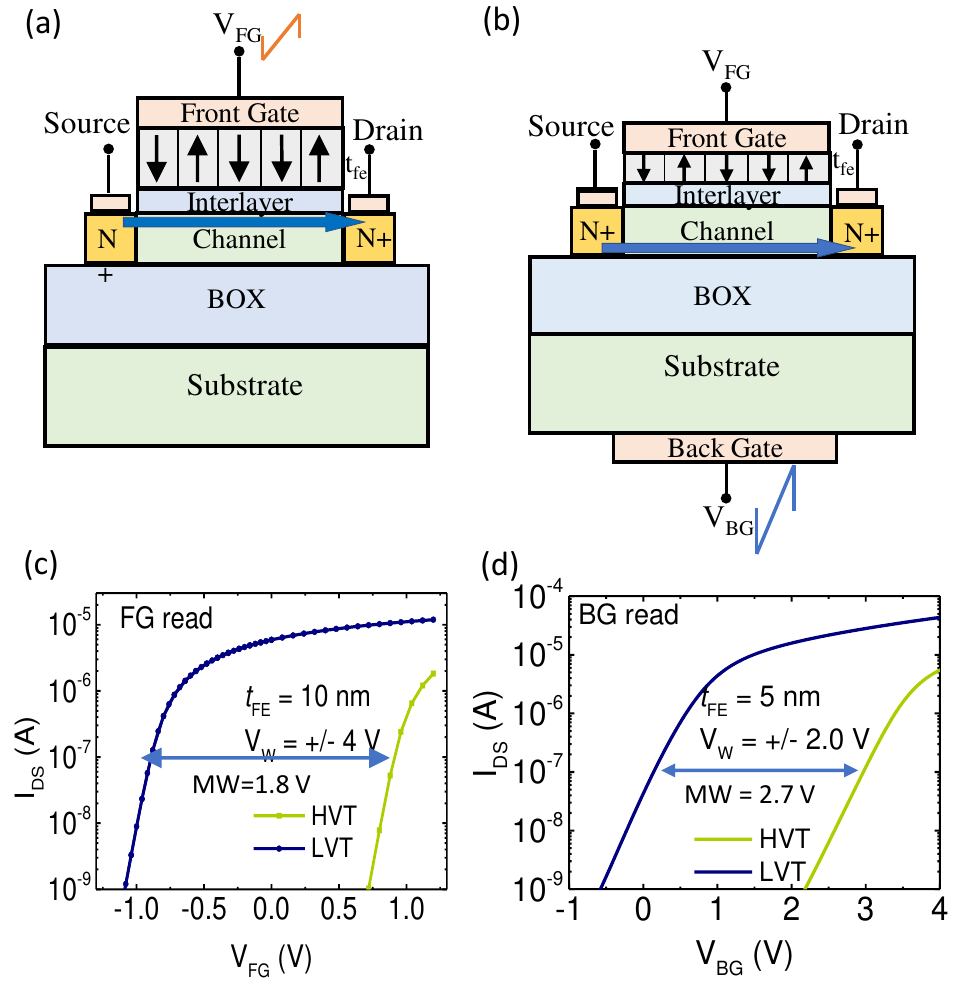}
    \vspace{-0.8ex}
    \caption{Schematic representation of FDSOI FeFET with (a) Single-gate and (b) Double-gate. Arrows in the channel region show the formation of the channel due to reading voltage at the front and back gates. I-V characteristic of (c) FG read FeFET with write voltage \vwrite = $\pm4$V and MW = 1.8V and (d) BG read FeFET with \vwrite = $\pm2$V and MW = 2.7V.}
    \vspace{-2ex}
    \label{fig:structure}
\end{figure}

A FeFET is composed of a FE layer integrated into the gate stack of a MOSFET. The coupling effect of FE capacitance and MOSFET gate capacitance exhibits a tunable hysteresis leading to the non-volatile feature. By applying a positive/negative voltage on the FG, the FE layer is polarized to place the underlying transistor in either the low-\vth (LVT) or high-\vth (HVT) state.

As the memory window (MW) is proportional to the FE thickness (\tfe), conventional SG-FeFETs need a thick FE layer (10nm) to achieve the required memory window (MW)~\cite{reliability_ted22,9939626}. However, the thick FE layer leads to severe charge trapping and requires a relatively high write voltage (+/-4V), which also hinders the technology scaling. Besides, as shown in Fig. \ref{fig:single_gate}, the write and read operations of SG-FeFET shares the common FG, hence frequent read operations can alter the polarization state causing accumulated read disturbance.




Recently, a double-gate FeFET structure is proposed to address the aforementioned issues~\cite{mulaosmanovic2021ferroelectric}. The device structure of DG-FeFET is similar to SG-FeFET except for the employment of BG, as shown in \cref{fig:double_gate}. For the SG-FeFET, the back side is the body bias (BB) and is not applied to write or read pulse. For the DG-FeFET, the write pulse and read pulse are applied to FG and BG separately, which can efficiently avoid the accumulated read disturbance. Additionally, the BG read scheme amplifies the MW of the DG-FeFET, as the comparison between in \cref{fig:SG_idvg} and \cref{fig:DG_idvg}. A 3nm \tfe is sufficient to achieve MW of  2.7V~\cite{reliability_ted22,mulaosmanovic2021ferroelectric}. Therefore, write voltage can be significantly reduced and endurance can be improved to $\!10^{10}$ level~\cite{tan2021ferroelectric}. A compact SPICE model for DG-FeFET is recently presented and is well calibrated with TCAD simulation results~\cite{kumar2022}.

However, two aspects need to be carefully considered for the DG-FeFET-based design. First, to support individual control for BG, isolated P-wells are required, resulting in area penalties. Second, though the BG read scheme can amplify the MW, it reduces the sub-threshold slope (SS) of DG-FeFET, as shown in Fig.~\ref{fig:DG_idvg}. In this case, the ON current of DG-FeFET is more sensitive to bias change. 


\subsection{Existing TCAM Designs}
\label{cam_design}

\begin{figure}[t]
    \begin{subfigure}{\columnwidth}
        \phantomsubcaption
        \label{fig:dg_example_macro}
        \phantomsubcaption
        \label{fig:sg_example_macro}
    \end{subfigure}
    \centerline{\includegraphics[trim=15 15 15 15, clip, width=1\linewidth] {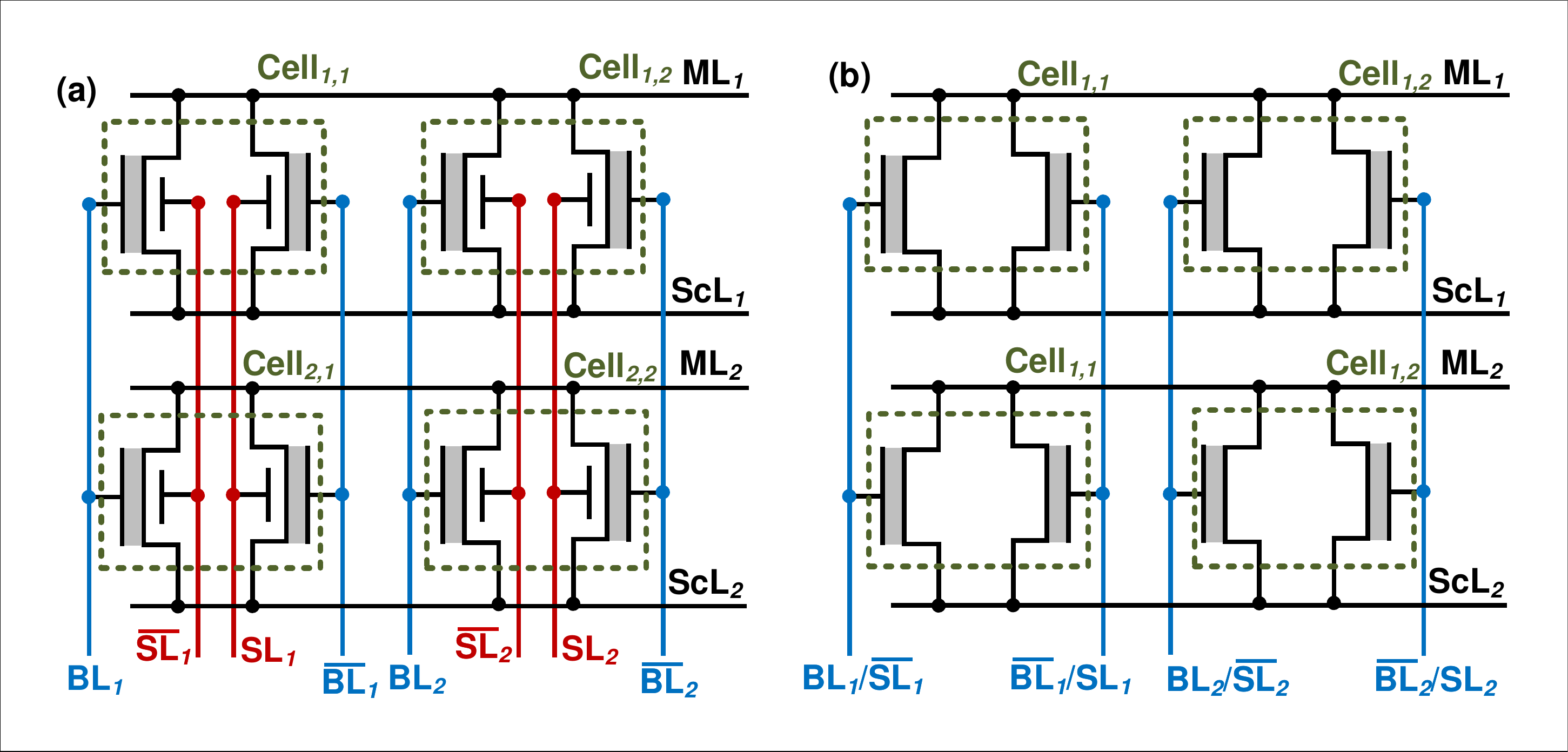}}
    \vspace{-3ex}
    \caption{The schematic comparison of $2\!\times\!2$ (a) 2DG-FeFET and (b) 2SG-FeFET TCAM.}
    \vspace{-5pt}
    \label{fig:2fefet}
\end{figure}

TCAM is the most widely used CAM type since TCAMs provide an additional `don't care' state (`X' state) to allow a wildcard operation in addition to the `0’ and `1’ states offered in the binary CAM (BCAM).
Conventional CMOS TCAM designs require 10-16 transistors per cell, thus the high energy and area costs limit their applications. NV-TCAMs are typically implemented by two NVM devices to encode `0' and `1' states using high and low resistance states with several control transistors, which are more compact and energy efficient. A 2T-2R compact TCAM structure is proposed based on either PCM~\cite{2t2pcm_jssc14} or RRAM~\cite{li_ted21}. A 2FeFET TCAM design is proposed to achieve an ultra-dense TCAM cell without any control transistors, which is the most widely-adopted FeFET TCAM design~\cite{yin_tcas18, ann_iccad20,searchd_tcad19,ni_ne19} (see Fig. \ref{fig:2fefet}(b)). 
The NV-TCAM designs based on two-terminal NVMs usually require current-driven write schemes and have a large leakage current for search. The low ON/OFF ratio and capacitance from the large access transistors also limit the search speed and word length of the CAM array. Conventional SG-FeFETs seem to be free from the issues of two-terminal NVMs, but the write voltage (+/-4V) is higher than most standard CMOS technologies hence posing challenges to the write drivers and high-voltage tolerance for other peripherals. 

\begin{figure}[t]
    \centering
    \includegraphics[trim=10 10 10 5, clip, width=0.75\linewidth] {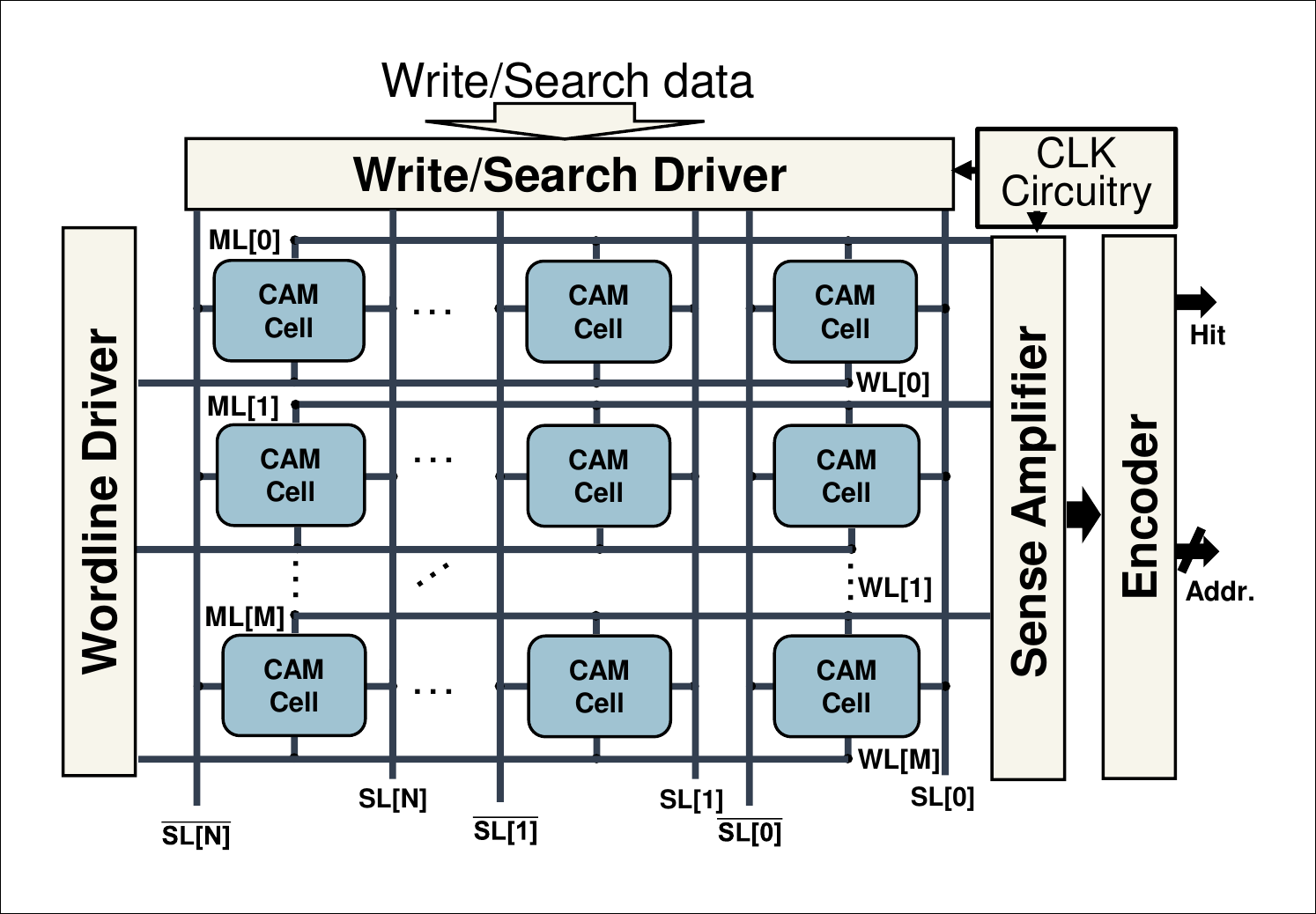}
    \vspace{-1.0ex}
    \caption{The architecture of a general \mTimesn NOR-type CAM array.}
    \vspace{-15pt}
    \label{fig:subarray}
\end{figure}

Besides, two NVMs per cell design is also relatively expensive and increases the overhead of peripherals, which is not only reflected in the cell-level metrics, it also impacts the array-level performance. Some NV-CAM designs are proposed to implement a TCAM cell with only one NVM device. A 3T1R RRAM TCAM~\cite{3t1rram_isscc15} and a similar 2.5T1R RRAM TCAM~\cite{2.5t1rram_isscc16} are proposed based on the voltage-divider concept, but they require complex control signals and additional access transistors. Recent work proposed a 2T1Fe CAM design, but it is only a BCAM design that cannot support a `don't care' state~\cite{2t1fe_dac22}. A 3T1Fe multi-bit CAM design is proposed, but the sensing circuit and control signals are complex~\cite{ramin_ted21}.
In a TCAM array, an input query is compared against all the stored entries in parallel and the address of the matched entry is returned. As shown in \cref{fig:subarray}, in a CAM array, match lines (MLs) are shared by the CAM cells in a row and sensed by the sense amplifier (SA). For a search operation, MLs are first precharged, and then the input query is applied to the search lines (SLs). If the corresponding ML discharges to 0, indicating a mismatch; otherwise, ML stays high, indicating a match. 


\section{DG-FeFET-Based TCAM Design}
\label{proposed_work}

In this section, we first discuss 2DG-FeFET TCAM design and point out its unique challenges. We then present our 1.5T1DG-Fe TCAM design and the shared driver architecture.

\subsection{2DG-FeFET TCAM}
\label{2fefet}

Fig. \ref{fig:2fefet}(b) depicts the 2DG-FeFET TCAM design. 
The operations of 2DG-FeFET TCAM are summarized in \cref{tab:2dg-fefet}. 
the bit lines (BLs) for the write operation and SLs for the search operation are separated and connected to the FG and BG of DG-FeFET, respectively. The 2DG-FeFET TCAM design inherits the advantages of DG-FeFET but also faces the corresponding challenges. First, for a \mTimesn TCAM array, $2N$ column-wise P-wells are required to accommodate the dedicated SLs for the $2N$ FeFETs in each column, which can be substantially expensive. Second, due to the reduced SS of the DG-FeFET, the search latency of 2DG-FeFET TCAM is longer than its SG-FeFET counterpart. Therefore, the straightforward 2DG-FeFET TCAM design may not be an ideal choice.



\begin{table}[t]
    \vspace{-1.5ex}
    \centering
    \caption{Operations of 2DG-FeFET TCAM Cell}
    \label{tab:2dg-fefet}
    \begin{tabular}{ccccccc}
    \hline
    \toprule
    \textbf{Operation} & \textbf{State} & \textbf{BL} & \textbf{\textoverline{BL}} & \textbf{SL} & \textbf{\textoverline{SL}}& \textbf{\begin{tabular}[c]{@{}c@{}}2FeFET states\end{tabular}} \\ 
    \midrule
    \multirow{3}{*}{\textbf{Write}}  & \textbf{0} & -\vwrite  & +\vwrite  & 0 & 0   & HVT / LVT  \\ 
                                     & \textbf{1} & +\vwrite  & -\vwrite  & 0 & 0   & LVT / HVT \\ 
                                     & \textbf{X} & -\vwrite  & -\vwrite  & 0 & 0   & HVT / HVT \\
    \midrule                                
    \multirow{2}{*}{\textbf{Search}} & \textbf{0} & 0 & 0 & \vs & 0 & \multirow{2}{*}{--} \\ 
                                     & \textbf{1} & 0 & 0 & 0   & \vs   &                   \\
    \bottomrule
    \vspace{-6pt} \\
    \multicolumn{7}{l}{\textbf{\vwrite = 2V;  \vs = 2V.}} \\
    \end{tabular}
    \vspace{-3ex}
\end{table}

\subsection{1.5T1DG-Fe TCAM}

\label{1.5t1fefet}
\begin{table}[t]
    \centering
    \caption{Operations of 1.5T1DG-Fe TCAM Cell}
    \label{tab:1.5t1dg-fefet}
    \begin{tabular}{ccccccc}
    \hline
    \toprule
    \textbf{Operation} & \textbf{State} & \textbf{BL} & \textbf{SeL} & \textbf{Wr/SL} & \textbf{SL} & \textbf{\begin{tabular}[c]{@{}c@{}}FeFET state\end{tabular}} \\ 
    \midrule
    \multirow{3}{*}{\textbf{Write}}  & \textbf{0} & -\vwrite   & 0  & \vdd & 0   & HVT (\roff)       \\ 
                                     & \textbf{1} & +\vwrite   & 0  & \vdd & 0   & LVT (\ron)         \\ 
                                     & \textbf{X} & \vm    & 0  & \vdd & 0   & MVT (\rmm)       \\
    \midrule                                
    \multirow{2}{*}{\textbf{Search}} & \textbf{0} & \vb          & \vsel & \vdd & \vdd & \multirow{2}{*}{--} \\ 
                                     & \textbf{1} & 0          & \vsel & 0   & 0   &                   \\
    \bottomrule
    \vspace{-6pt} \\
    \multicolumn{7}{l}{\textbf{\vwrite = 2V; \vm =1.6V; \vsel = 2V; \vdd =0.8V; \vb =0.25V. }} \\
    \end{tabular}
    \vspace{-4ex}
\end{table}

\begin{figure}[t]
\begin{center}
\centerline{\includegraphics[trim=15 15 15 15, clip, width=0.8\linewidth] {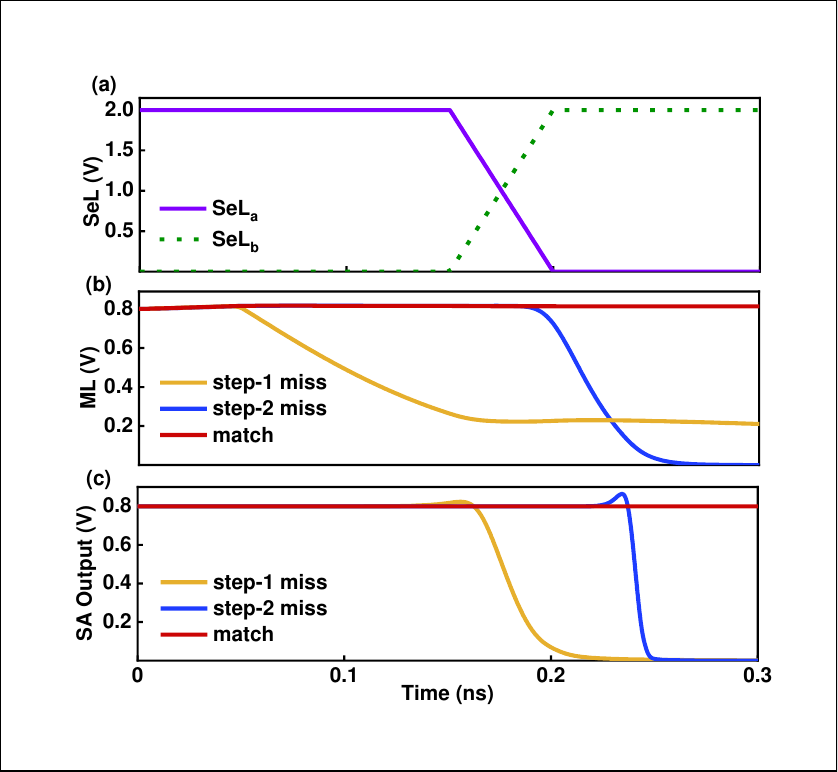}}
\vspace{-1.0ex}
\caption{The transient waveform of (a) select signals \selA and \selB, and (b) ML and (c) SA output of step-1 miss, step-2 miss, and match cases.}
\vspace{-6ex}
\label{fig:wv}
\end{center}
\end{figure}

\begin{figure*}[t]
    \begin{subfigure}{\columnwidth}
        \phantomsubcaption
        \label{fig:proposed_cell}
        \phantomsubcaption
        \label{fig:equivalent_write_read_circuit}
        \phantomsubcaption
        \label{fig:dg_proposed_macro}
        \phantomsubcaption
        \label{fig:sg_proposed_macro}
    \end{subfigure}
    \begin{center}
    \centerline{\includegraphics[trim=15 15 15 15, clip, width=1\linewidth] {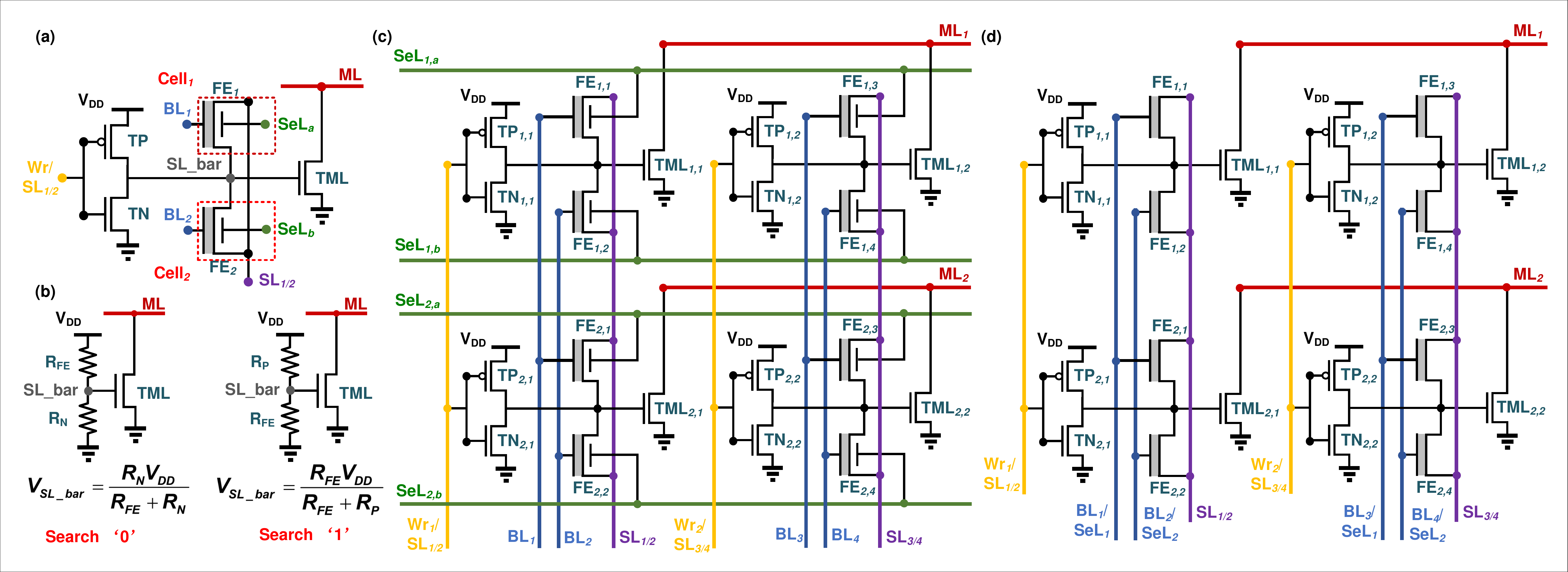}}
    \vspace{-1.0ex}
    \caption{(a) The proposed 1.5T1DG-Fe TCAM in 2-cell pair. (b) The equivalent circuit of searching `0' and searching `1'. The $2\!\times\!4$ (c) 1.5T1DG-Fe TCAM array and (d) 1.5T1SG-Fe TCAM array.}
    \vspace{-6ex}
    \label{fig:array}
    \end{center}
\end{figure*}

\subsubsection{\textbf{1.5T1DG-Fe TCAM Overview}}
\label{sec:overview}

To address the issues of the 2DG-FeFET TCAM design, we propose a 1.5T1DG-Fe TCAM design that only uses a single FeFET in a TCAM cell, as shown in \cref{fig:proposed_cell}. Based on the voltage-divider scheme, we use the HVT, LVT, and medium-\vth (MVT) of a DG-FeFET to encode the ternary states of TCAM. To reduce the cell area, every two DG-FeFETs are grouped in a 2-cell pair and adopt a two-step search to share the control transistors TP, TN, and TML. TML, a small NMOS transistor, is connected to the ML for every two TCAM cells, which reduces the search latency and ML precharge energy. The DG-FeFET devices do not directly participate in the ML discharge process, hence the impact of reduced SS is mitigated. Compared with the 2DG-FeFET TCAM design, for every two TCAM cells, the direct capacitance load on the ML reduces from 4 large DG-FeFET devices to 1 small NMOS transistor, leading to a shorter ML delay.

\subsubsection{\textbf{TCAM Cell Operation}}
\label{sec:write+search}

The operations of the 1.5T1DG-Fe TCAM design are shown in \cref{tab:1.5t1dg-fefet}. During the write operation, to keep the source, drain, and BG of the DG-FeFET to the ground level, the Wr/SL is set to \vdd, and SL and SeL are set to 0. The write voltage is applied on the BL to program the DG-FeFET state. Besides the `0' (HVT, \roff)  and `1' state (LVT, \ron), an `X' state is required, and the corresponding DG-FeFET resistance \rmm is between \roff and \ron. The array-level write scheme will be discussed in \cref{array+reliability}. 

During the search operation, TCAM cell$_1$ and cell$_2$ are searched in two steps. Select signals \selA and \selB are connected to the BG of cell$_1$ and cell$_2$, respectively. The select voltage \vsel equals the BG read voltage of the DG-FeFET. Cell$_1$ is searched in the first step (\selA = \vsel, \selB = 0), and cell$_2$ is searched in the second step (\selA = 0, \selB = \vsel). We use cell$_1$ as an example to explain the way to apply the search query, and cell$_2$ follows the same process. 

To ensure the correct operation of the TCAM, the resistance values of TN, TP, and DG-FeFET must be carefully selected. Specifically, The ON resistance of TP (\rp), TN (\rn) and the DG-FeFET resistance of `0' (\rfe = \roff), `1' (\rfe = \ron), and `X' (\rfe = \rmm) states should satisfy: 
\begin{equation} \label{equ:res}
\textbf{\ron $<$ \rn $<$ \rmm $<$ \rp $\ll$ \roff}
\end{equation}
When searching for `0', \vdd is applied to Wr/SL and SL, thus TN is turned on and the equivalent circuit is shown in \cref{fig:equivalent_write_read_circuit}. The voltage of \slBar can be estimated as
\begin{equation} \label{equ:search0}
    \vslBar = \frac{\vdd \times \rn}{\rfe + \rn}
\end{equation}
If the stored value is `0' (\rfe = \roff), \rfe is much larger than \rn, which keeps the voltage of \slBar (\vslBar) smaller than the threshold voltage of TML (\vth). TML is turned off and the ML stays high, resulting in a match. If the stored value is `1' (\rfe = \ron), \rfe is smaller than \rn, and \vslBar is higher than \vth. TML is turned on and the ML discharges through TML, resulting in a mismatch. To keep the \ron relatively constant when connecting in series with \rn, a small bias (\vb) is applied to the BL to provide better FG to source 
voltage potential of DG-FeFET.

When searching for `1', Wr/SL and SL are connected to the ground, thus TP is turned on and the equivalent circuit is shown in \cref{fig:dg_proposed_macro}. The voltage of \slBar can be estimated as
\begin{equation} \label{equ:search1}
    \vslBar = \frac{\vdd \times \rfe}{\rfe + \rp}
\end{equation} 
Similarly, if the stored value is `0' (\rfe = \roff), \rfe is much larger than \rp, which makes \vslBar higher than \vth. TML is turned on and the ML discharges, resulting in a mismatch. If the store value is `1' (\rfe = \ron), \rfe is smaller than \rp, hence \vslBar is lower than the \vth of TML. TML is turned off and the ML stays high, resulting in a match.
For the TCAM cell storing the `X' state, the corresponding \rmm is between \rn and \rp, hence regardless of searching `0' or `1', \vslBar is always below \vth, and TML is always turned off, achieving the `don't care' function.

\subsubsection{\textbf{TCAM Array \& Early Search Termination}} 
\label{array+reliability}

A $2\!\times\!4$ array of the proposed 1.5T1DG-Fe TCAM is shown in \cref{fig:dg_proposed_macro}. Signals Wrs/SLs, BLs, and SLs are shared column-wise, and \selA, \selB, and MLs are shared row-wise. Since the search controls (i.e., \selA/\selB) are connected row-wise in our 1.5T1DG-Fe TCAM design, $2M$ separated P-wells are required for a \mTimesn TCAM array. Compared to the $2N$ P-wells required by the 2DG-FeFET design, though the number of P-wells is comparable if $M$ and $N$ are similar, the total number of DG-FeFETs in the TCAM array is reduced by half. 



We should point out that our proposed 1.5T1DG-Fe TCAM uses three \vth levels hence three-step write is required. Though one more write step than the 2FeFET TCAM design is needed to write the `X' state, write is much less frequent than search. Other NV-TCAM designs based on single NVMs have similar 3-step write schemes~\cite{2.5t1rram_isscc16,3t1rram_isscc15}. 



At the array level, a two-step search with early termination is adopted for the 1.5T1DG-Fe TCAM, as shown in Fig. \ref{fig:wv}. In the first step, the search query is applied to search all the cell$_1$s (\selA = \vsel) in the 2-cell pair, as shown in Fig. \ref{fig:wv}(a). If no mismatch exists and the ML stays high, then we search all the cell$_2$s in the second step (\selB = \vsel). If both two steps find no mismatch, ML stays high, and SA outputs `1', indicating a match for the entry (match case in Fig. \ref{fig:wv}(b) and (c)); otherwise, ML discharges to the ground and SA outputs `0', indicating a mismatch (step-2 miss case in Fig. \ref{fig:wv}(b) and (c)). If the stored entries are mismatched in the first step, the search operation terminates, then ML discharges to the ground and SA outputs `0', indicating a mismatch (step-1 miss case in Fig. \ref{fig:wv}(b) and (c)). The \selB signal (green dot line in Fig. \ref{fig:search_eva}) is grounded and will not apply \vsel. In this case, the DG-FeFET is in the OFF state, hence the leakage current going through the voltage divider structure can be reduced (see Fig. \ref{fig:array}(b)), leading to search energy savings. In real-world applications, most of the stored entries return mismatches. Therefore, the energy saving from the early termination is significant. In addition, regardless of the match results of the first and second steps, the ML is precharged once, further improving energy efficiency.

\subsubsection{\textbf{High Voltage Driver Optimization}}
\label{driver}


\begin{figure}[t]
\begin{center}
\centerline{\includegraphics[trim=15 15 15 15, clip, width=1\linewidth] {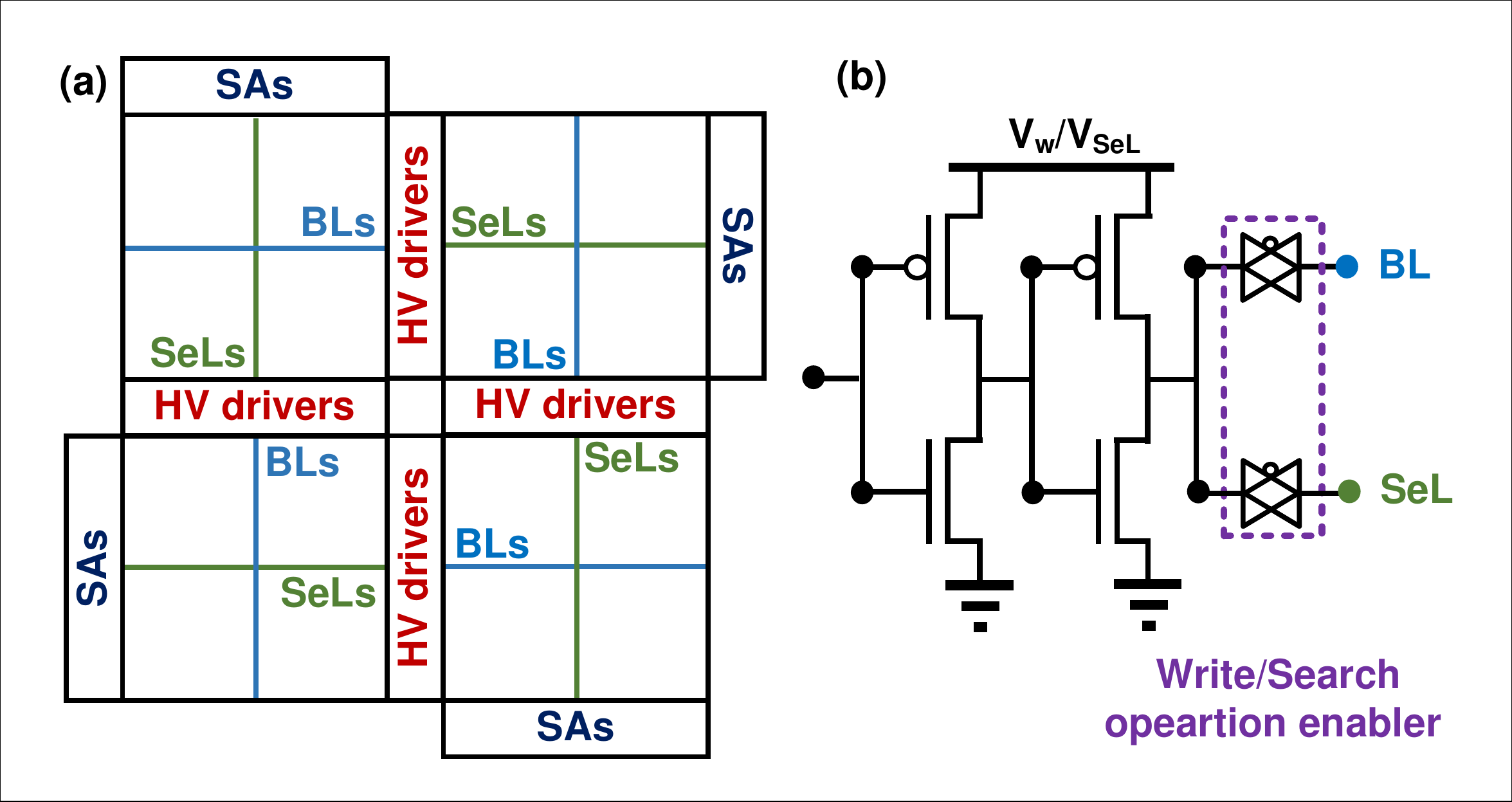}}
\caption{(a) The shared HV drivers architecture between the adjacent TCAM subarrays, and (b) the schematic of a shared HV driver. }
\vspace{-7ex}
\label{fig:driver}
\end{center}
\end{figure}

If the search and write voltage of an NV-TCAM is the same. they would be able to share the same driver, which can save not only the area but also increase the driver utilization and reduce the leakage power. 

Toward this end, we have explored the DG-FeFET design space and identified a delicately optimized combination of device parameters like gate work-function tuning such that the LVT write voltage and read voltage of the DG-FeFET are the same voltage level (2.0V) and the resulting Id-Vg curve is shown in Fig.\ref{fig:DG_idvg}. The DG-FeFET can achieve 2.7V MW and $10^4$ level ON/OFF ratio.


Given that the search and LVT programming share the same voltage level, we design a specific driver-sharing scheme for the 1.5T1DG-Fe TCAM design, shown in Fig. \ref{fig:driver}(a). The shared drivers drive the BLs in the write operation and drive the SeLs in the search operation. Since the BLs and SeLs are placed perpendicularly in each subarray and they are not employed at the same time, we share the HV driver between adjacent subarrays in a time-multiplexed manner. The adjacent subarray is rotated by 90\degree \xspace  and four subarrays compose a mat. The number of drivers is cut in half, which not only reduces the driver area but also increases driver utilization. The schematic of a shared HV driver is shown in Fig. \ref{fig:driver}(b), and the driver connection is controlled by the write/search enabled signal. 

\section{1.5T1SG-Fe TCAM}

\begin{table}[t]
\centering
\caption{Operations of 1.5T1SG-Fe TCAM Cell}
\label{tab:1.5t1sg-fefet}
\begin{tabular}{cccccc}
\hline
\toprule
\textbf{Operation} & \textbf{State} & \textbf{BL/SeL} &  \textbf{Wr/SL} & \textbf{SL} & \textbf{\begin{tabular}[c]{@{}c@{}}FeFET state\end{tabular}} \\ 
\midrule
\multirow{3}{*}{\textbf{Write}}  & \textbf{0} & -\vwrite    & \vdd & 0   & HVT (\roff)       \\ 
                                 & \textbf{1} & +\vwrite    & \vdd & 0   & LVT (\ron)         \\ 
                                 & \textbf{X} & \vm      & \vdd & 0   & MVT (\rmm)       \\
\midrule                                
\multirow{2}{*}{\textbf{Search}} & \textbf{0} & \vsel        & \vdd & \vdd & \multirow{2}{*}{--} \\ 
                                 & \textbf{1} & \vsel        & 0   & 0   &                   \\
\bottomrule
\vspace{-6pt} \\
\multicolumn{6}{l}{\textbf{\vwrite = 4V; \vm = 3.2V; \vsel =0.8V; \vdd =0.8V. }} \\
\end{tabular}
\vspace{-2ex}
\end{table}

The proposed 1.5T1DG-Fe TCAM design can also be adapted to using SG-FeFETs. Compared with 1.5T1DG-Fe TCAM, the select signal and write signal (BL/SeL) are merged and connected to the FG of SG-FeFET, as shown in \cref{fig:sg_proposed_macro}. The operations are summarized in \cref{tab:1.5t1sg-fefet}. The 1.5T1SG-Fe TCAM design has a smaller cell area compared with the DG-FeFET-based counterpart because the SG-FeFET-based design does not require individual BG control. 


\section{Evaluation}
\label{evaluation}

\begin{table*}[t]
\centering
\caption{FoM comparison of CMOS and FeFET TCAM designs}
\label{tab:foms}
\begin{tabular}{cccccc}
\hline
\toprule
\textbf{FoM}                 &  \textbf{16T CMOS$\dagger$ \cite{16t_esscirc15}} & \textbf{2SG-FeFET} &  \textbf{2DG-FeFET} & \textbf{1.5T1SG-Fe}   & \textbf{1.5T1DG-Fe}   \\ \hline
\textbf{Write voltage} & 0.9V & $\pm4$V & $\pm2$V & $\pm4$V, 3.2V & $\pm2$V, 1.6V \\
 \rowcolor{Gray}
\textbf{FE thickness} & N.A. & 10nm &  5nm & 10nm & 5nm \\
\textbf{Cell area ($um^2$)} & 0.286  ($1\times$) & 0.095 ($3.01\times$)  & 0.204 ($1.40\times$) & 0.108 ($2.65\times$) & 0.156 ($1.83\times$)  \\
 \rowcolor{Gray}
\textbf{Write energy/cell} (fJ)  & N.A. & 1.63 ($1\times$) & 0.81 ($2\times$) & 0.82 ($2\times$)& 0.41 ($4\times$) \\
   &  &  &  & 1 step: 159  & 1 step: 231            \\
\multirow{-2}{*}{\textbf{Search Latency} (ps)} & \multirow{-2}{*}{235 ($1\times$)}  &\multirow{-2}{*}{582 ($0.4\times$)} &  \multirow{-2}{*}{1147 ($3.01\times$)} & 2 steps: 351 ($0.67\times$) & 2 steps: 481 ($0.49\times$)\\ 
\rowcolor{Gray}
 &  &  &  & 1 step: 0.11 & 1 step: 0.13 \\
 \rowcolor{Gray}
 &  &  &  & 2 steps: 0.16 & 2 steps: 0.21 \\
 \rowcolor{Gray}
 \multirow{-3}{*}{\textbf{Search Energy/cell} (fJ)} & \multirow{-3}{*}{ 0.53 ($1\times$)}  & \multirow{-3}{*}{0.17 ($3.12\times$)} &  \multirow{-3}{*}{0.25 ($2.12\times$)} & Average$*$: 0.12 ($4.42\times$) & Average$*$: 0.14 ($3.79\times$) \\    
\bottomrule  
\multicolumn{6}{l}{$*$ The average search energy consumption per cell in real-world applications, assuming 90\% step-1 miss rate. } \\
\multicolumn{6}{l}{$\dagger$ The 16T CMOS TCAM is implemented in 14nm SOI technology with 64-bit word length, and data is from simulated results. } \\
\end{tabular}
\vspace{-2ex}
\end{table*}

\begin{figure}[t]
    \begin{center}
    \centerline{\includegraphics[trim=20 15 15 20, clip, width=1.0\linewidth] {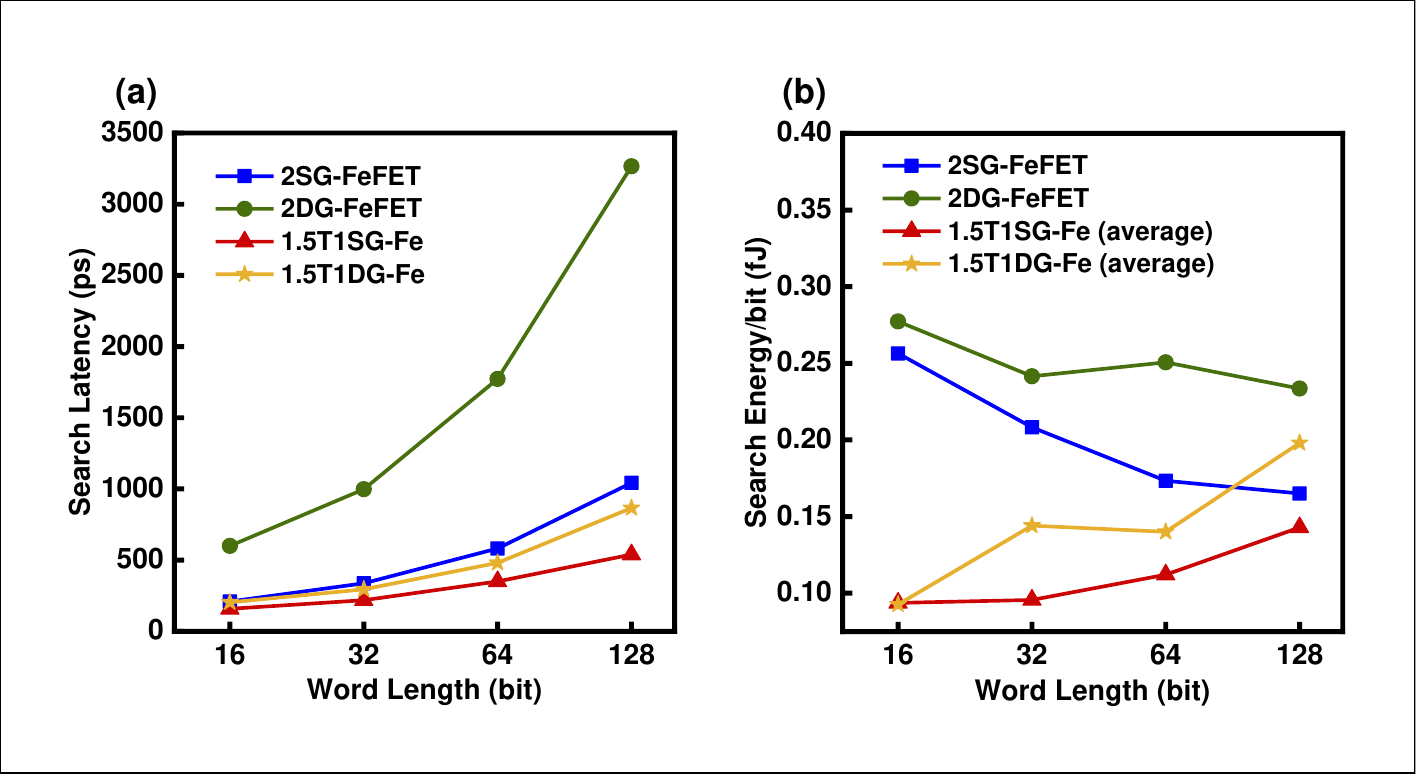}}
    \vspace{-1.5ex}
    \caption{Word length impact on (a) search latency and (b) search energy for 2SG-FeFET,  2DG-FeFET,  1.5T1SG-Fe, and 1.5T1DG-Fe TCAM designs.}
    \label{fig:search_eva}
    \vspace{-6ex}
    \end{center}
\end{figure}

\subsection{Simulation Setup}
\label{sec:setup}

To evaluate our proposed 1.5T1Fe TCAM and benchmark with 2FeFET TCAM design, we have performed extensive SPICE simulations. A 14nm BSIM-IMG model is calibrated with the experimental data~\cite{fdsoi} and is used for the MOSFETs modeling. To ensure a fair comparison, we use the 14nm FDSOI SG-FeFET and DG-FeFET models based on the same 14nm BSIM-IMG model and are calibrated with TCAD simulation results~\cite{kumar2022}. The device size of SG-FeFETs and DG-FeFETs is $20 \times 50$ nm. The FE layer thickness of SG-FeFET and DG-FeFET is 10nm and 5nm, respectively. The wire parasitics are extracted from Eva-CAM~\cite{eva-cam}.

\subsection{TCAM Array Evaluation}
\label{sec:evaluation}

Tab.~\ref{tab:foms} compares the figures-of-merit (FoM) of four FeFET TCAM designs with standard 16T CMOS TCAM designs. We comprehensively evaluate the four designs from both FeFET devices and CAM cell circuits aspects. The FoM includes write voltage, FE thickness of FeFET, TCAM cell area, write energy per cell,  search latency, and search energy per cell. The data of the four designs are obtained from the simulation results of a 64$\times$64 TCAM array size. A 16T CMOS TCAM design implemented in 14nm SOI technology with 64-bit word length is also included to fairly compare with mature TCAM technology~\cite{16t_esscirc15}. The FoM of 2SG-FeFET design has been extensively compared with other NV-TCAM designs \cite{yin_tcas18,1t2fe_date21,2t1fe_dac22}, hence here we mainly focus on the FeFET TCAM evaluation and set the 16T CMOS design as the baseline.   
For the DG-FeFET-based design, the write voltage is reduced to +/-2V, which is only half of the SG-FeFET-based design. The reduced write voltage can not only improve the endurance but also lead to greater advantages from the peripherals that may not be fully reflected in cell-level metrics. 

The TCAM cell areas are estimated by their layouts based on~\cite{cand_tcas22}. We consider the large spacing between different P-wells in our estimation. Due to the separated P-wells for BG control, the cell area of the DG-FeFET-based design is larger than the SG-FeFET-based counterpart. Besides, to achieve ideal resistance relations presented in equation (\ref{equ:res}), relatively large TP and TN transistors are required for the 1.5T1Fe TCAM design, so the cell area of our proposed 1.5T1SG-Fe and 1.5T1DG-Fe designs are larger than the 2SG-FeFET designs. But the cell areas of all four FeFET TCAM designs are smaller than 16T CMOS TCAM in the same 14nm technology node. 
The 2FeFET design encodes TCAM states using HVT and LVT of 2 FeFETs in a complimentary manner, so the write energy per cell does not depend on the stored state of TCAM. Our proposed 1.5T1Fe design only uses one FeFET to store the TCAM state, hence stored data impacts the write energy evaluation. To do a fair comparison, we evaluate the average case that half of the 1.5T1DG-Fe cells are programmed to `0' and half of the cells are programmed to `1'. DG-FeFET designs have lower write energy than their SG-FeFET counterpart for their lower write voltage. In addition, due to the single FeFET-based design, the write energy per cell of 1.5T1Fe TCAM is reduced to half of the 2FeFET design. As a result, the implemented 2DG-FeFET, 1.5T1SG-Fe, and 1.5T1DG-Fe designs achieve 2$\times$, 2$\times$, and 4$\times$ write energy compared to the 2SG-FeFET design, respectively.  
The search latency is impacted by the total capacitance and resistance on the ML. We consider the worst-case latency of one-cell mismatch. For the proposed 1.5T1Fe TCAM design, every two cells only have a minimum size NMOS transistor connected to the ML, thus the search latency is reduced significantly. We include the latency of the two-step search and leave some time slack for the search signal switching between the two steps. The 1.5T1SG-Fe and 1.5T1DG-Fe TCAM designs achieve 1.66$\times$ and 1.21$\times$ search latency improvement compared with the 2SG-FeFET design. The search latency of the 1.5T1DG-Fe design is slightly slower than the 1.5T1SG-Fe design, because of the higher ON resistance of DG-FeFET. For the store `1' search `0' case, as discussed in Sec. \ref{sec:write+search}, the TML is not fully turned on hence limiting the ML discharge speed. The 2DG-FeFET design search latency is longer than the 2SG-FeFET design for a similar reason. 
The 16T CMOS search latency from simulated results \cite{16t_esscirc15} is shorter than our 1.5T1Fe TCAM design, which may be due to CMOS TCAM using the minimum channel size to achieve a faster switch in the advanced technology node. 
 
The search energy of the TCAM cell mainly consists of the ML precharge, SA, and the search signals energy consumption. For the 2FeFET design, the ML precharge and SA energy consume the majority of the search energy. For the 1.5T1Fe TCAM design, the ML precharge energy is smaller than the 2FeFET design, but the voltage divider structure produces a non-negligible current during the search operation, especially for the FeFETs in LVT states (\rfe = \ron), leading to higher energy consumption for the search signals. We evaluate the average case in that half of the cells store `0' and half of the cells store `1'. Additionally, the early search termination scheme helps to save considerable search energy. We report both the single-step search energy and the total two-step search energy in Tab.~\ref{tab:foms}. In real-world applications, typically more than 95\% stored entries return mismatches in the first step but this depends on the search pattern~\cite{2.5t1rram_isscc16}. Here we assume 90\% 1-step mismatch rates for a pessimistic estimation and calculate the average search energy per cell. The proposed 1.5T1SG-Fe and 1.5T1DG-Fe TCAM designs achieve 1.42$\times$/1.21$\times$ and 4.42$\times$/3.79$\times$ search energy improvement compared to the 2SG-FeFET and 16T CMOS design, respectively. Due to the longer sensing time of the 2DG-FeFET and 1.5T1DG-Fe designs, their search energy is higher than the SG-FeFET counterpart.

\subsection{Design Space Exploration}
\label{sec:dse}

Fig. \ref{fig:search_eva}(a) and (b) show the word length impact on the search latency and energy of the four SG-FeFET and DG-FeFET TCAM designs. The search latency and energy evaluation methods are consistent with Sec. \ref{sec:evaluation}. As the word length increases, the associated ML capacitance increase, hence the search latency of the four TCAM designs increases accordingly. But the latency increase trends of the 1.5T1Fe design are slower than the 2FeFET design, which shows better performance scalability. The search energy per cell shows different trends of the 2FeFET and 1.5T1Fe TCAM designs. For the 2FeFET design, the search energy per cell decreases with word length increases, because the increased number of CAM cells per word amortizes the energy consumption of the SA. However, for the 1.5T1Fe TCAM designs, as the word length increases, the search latency increases, hence the voltage divider structure dominates the total energy consumption, which suppresses the energy amortization effects of SA.


\section*{Conclusion}
\label{conclusion}

This paper introduces a novel 1.5T1Fe TCAM design based on DG-FeFETs that employs only one DG-FeFET per TCAM cell. We systematically analyze the advantages and challenges of DG-FeFETs design. The 1.5T1DG-Fe TCAM design alleviates the BG control overhead and achieves fast parallel search with high energy efficiency. A search early termination scheme is proposed to further reduce the search energy. A shared driver design is presented with device-circuit co-optimization to reduce the peripherals overhead. The proposed 1.5T1Fe TCAM can also be implemented by conventional SG-FeFET achieving superior search speed and energy efficiency compared to 2SG-FeFET.

\section*{Acknowledgement}
\label{ack}
    
This research was supported in part by the Semiconductor Research Corporation (SRC) Logic and Memory Devices Program (LMD), and by AI Chip Center for Emerging Smart Systems (ACCESS) sponsored by InnoHK funding, Hong Kong SAR.

\bibliographystyle{IEEEtran.bst}
\bibliography{bib_short.bib}

\end{document}

%% file: glossar.tex
\newglossaryentry{vdd}{
	name={\ensuremath{\text{V}_\text{DD}}},
	description={supply voltage},
	symbol=Vdd,
	sort=vdd}
\newglossaryentry{vth}{
	name={\textit{V}\textsubscript{TH}},
	description={threshold voltage},
	symbol=Vth,
	sort=vth}
\newglossaryentry{gnd}{
	name={GND},
	description={El. ground},
	symbol=GND}

\newglossaryentry{ntype}{
	name={n-type},
	description={pnp doted transistor. Other known names are nmos or nfet}}
\newglossaryentry{ptype}{
	name={p-type},
	description={npn doted transistor. Other known names are pmos or pfet}}

\newglossaryentry{vds}{
	name={\ensuremath{\text{V}_\text{DS}}},
	description={Voltage difference between drain and source contact of a transistor},
	sort=vds}
\newglossaryentry{mwbg}{
    name=MW\textsubscript{BG},
    description= MWBG, symbol=mwbg}
\newglossaryentry{mwfg}{
    name=MW\textsubscript{FG},
    description= MWFG, symbol=mwfg}
\newglossaryentry{gammabf}{
    name=\ensuremath{\gamma\textsubscript{BF}},
    description= gammaBF, symbol=gammabf}
\newglossaryentry{qfe}{
    name=Q\textsubscript{FE},
    description= QFE, symbol=qfe} 
\newglossaryentry{vfe}{
    name=V\textsubscript{FE},
    description= VFE, symbol=vfe}

\newglossaryentry{wport}{
    name=WP,
    description= write port , symbol=wport}

\newglossaryentry{rport}{
    name=RP,
    description= read port , symbol=rport}

\newglossaryentry{wpbar}{
    name=$\overline{\text{WP}}$,
    description= write port bar , symbol=wpbar}

\newglossaryentry{rpbar}{
    name=$\overline{\text{RP}}$,
    description= read port bar, symbol=rpbar}
    
\newglossaryentry{qfix}{
    name=\textit{Q}\textsubscript{FIX},
    description= Fixed charge at interface, symbol=qfix}
\newglossaryentry{qfixp}{
    name=\textit{Q}\textsubscript{FIX+},
    description= Fixed charge at interface, symbol=qfix+}
\newglossaryentry{qfixn}{
    name=\textit{Q}\textsubscript{FIX-},
    description= Fixed charge at interface, symbol=qfix-}
    
\newglossaryentry{pfe}{
    name=\textit{P}\textsubscript{FE},
    description= Polarization in the ferroelectric layer,symbol=pfe}	
\newglossaryentry{pfep}{
    name=\textit{P}\textsubscript{FE+},
    description= up wards polarization in the ferroelectric  layer,symbol=pfep}
    
\newglossaryentry{pfen}{
    name=\textit{P}\textsubscript{FE-},
    description= down wards polarization in the ferroelectric  layer,symbol=pfen}

\newglossaryentry{tfe}{
    name=\textit{t}\textsubscript{FE},
    description= ferroelectric thickness,symbol=tfe}  

\newglossaryentry{bodyFactor}{
    name={\ensuremath{\gamma_{\text{BF}}}},
    description= body factor,
    symbol=gammaBF}
    
\newglossaryentry{vfg}{
    name={\ensuremath{\text{V}_{\text{FG}}}},
    description= Front gate voltage,
    symbol=vfg}
\newglossaryentry{vbg}{
    name={\ensuremath{V_{\text{BG}}}},
    description= Back gate voltage,
    symbol=vbg}
    
\newglossaryentry{vaux}{
    name={\ensuremath{V_{\text{aux}}}},
    description= aux voltage,
    symbol=vaux}
    
\newglossaryentry{vin}{
    name={\ensuremath{V_{\text{in}}}},
    description= in voltage,
    symbol=vin}   
    
\newglossaryentry{tauv}{
    name={\ensuremath{\tau_{\text{v}}}},
    description= in voltage,
    symbol=tauv}  
    
\newglossaryentry{paux}{
    name={\ensuremath{P_{\text{aux}}}},
    description= aux polarization,
    symbol=paux}   
    
\newglossaryentry{poff}{
    name={\ensuremath{P_{\text{off}}}},
    description= off polarization,
    symbol=poff}  
    
\newglossaryentry{mu}{
    name={\ensuremath{m_{\text{u}}}},
    description= upwards slope,
    symbol=poff}      

\newglossaryentry{vdel}{
    name={\ensuremath{V_{\text{del}}}},
    description= aux polarization,
    symbol=vdel}

\newacronym{nvm}{NVM}{Non-Volatile Memory}
\newacronym{hd}{HD}{Hamming distance}

\newacronym{fet}{FET}{Field Effect Transistor}
\newacronym{mosfet}{MOSFET}{Metal Oxide Semiconductor \acrlong{fet}}
\newacronym{fefet}{FeFET}{Ferroelectric \gls{fet}}
\newacronym{ram}{RAM}{Random Access Memory}
\newacronym{cmos}{CMOS}{Complementary MOS}

\newacronym{fe}{FE}{ferroelectric}

\newacronym{stt}{STT-MRAM}{Spin-Transfer Torque Magnetic RAM}
\newacronym{rram}{ReRAM}{Resistive RAM}

\newacronym{finfet}{FinFET}{Fin \acrlong{fet}}

\glsunset{fet}
\glsunset{finfet}

\newcommand{\selA}{\ensuremath{\text{SeL}_\text{a}}\xspace}
\newcommand{\selB}{\ensuremath{\text{SeL}_\text{b}}\xspace}
\newcommand{\slBar}{\ensuremath{\text{SL\_bar}}\xspace}

\newcommand{\vwrite}{\ensuremath{\text{V}_\text{w}}\xspace}
\newcommand{\vm}{\ensuremath{\text{V}_\text{m}}\xspace}
\newcommand{\vb}{\ensuremath{\text{V}_\text{b}}\xspace}
\newcommand{\vs}{\ensuremath{\text{V}_\text{s}}\xspace}
\newcommand{\vsel}{\ensuremath{\text{V}_\text{SeL}}\xspace}
\newcommand{\vslBar}{\ensuremath{\text{V}_\text{SL\_bar}}\xspace}

\newcommand{\ron}{\ensuremath{\text{R}_\text{ON}}\xspace}
\newcommand{\roff}{\ensuremath{\text{R}_\text{OFF}}\xspace}
\newcommand{\rmm}{\ensuremath{\text{R}_\text{M}}\xspace}
\newcommand{\rp}{\ensuremath{\text{R}_\text{P}}\xspace}
\newcommand{\rn}{\ensuremath{\text{R}_\text{N}}\xspace}
\newcommand{\rfe}{\ensuremath{\text{R}_\text{FE}}\xspace}

\newcommand{\mTimesn}{\ensuremath{M\!\times\!N}\xspace}

\newcommand{\vdd}{\gls{vdd}\xspace}

\newcommand{\vth}{\gls{vth}\xspace}

\newcommand{\tfe}{\gls{tfe}\xspace}